\documentclass{PoS}

\usepackage{epsfig}

\title{Recent Experimental Results on Soft Strong Interactions}

\ShortTitle{Soft Strong Interactions}

\author{\speaker{Paul Newman} \\
        School of Physics \& Astronomy, 
The University of Birmingham, B15 2TT, UK. \\
        E-mail: \email{p.r.newman@bham.ac.uk}}

\abstract{This report contains selected recent highlights
from experimental studies of soft and semi-hard strong interactions from the Large
Hadron Collider, HERA and the Tevatron. The subject
is broken down into diffractive and non-diffractive channels and the
most significant recent progress and open questions in each area are summarised.}

\FullConference{36th International Conference on High Energy Physics,\\
		July 4-11, 2012\\
		Melbourne, Australia}

\begin{document}

\section{Introduction}

The Melbourne ICHEP 2012 conference will remain in the memories of
all who attended, as the meeting where a 
Higgs boson discovery was first announced and where information on its
properties began to emerge. Although it is also concerned with the study of the new 
energy-frontier physics landscape
offered by the Large Hadron Collider (LHC), the material covered in this contribution
could hardly be more different: whereas a Standard Model Higgs boson 
with a mass of around $125 \ {\rm GeV}$ is produced once in around $10^{10}$
$pp$ collisions, soft strong interactions are at work in almost
all LHC collisions. 
Paradoxically, our understanding of the mechanisms at work in these most
ordinary of processes is much poorer than is the case for Higgs production. 
Whilst availability of data is not an issue,
the complexity of the interactions involved and the lack of a precision theory
in the absence of a large momentum or energy scale make soft strong interactions
far less predictable and more poorly understood than the physics of the Higgs
sector. The study of the bulk of the LHC cross section under what are often
referred to as `minimum bias' conditions is therefore in a very real sense,
virgin `swagman in billabong' \cite{waltzing} territory.
In addition to the new measurements
from the LHC, our models of soft strong interactions
have also recently been constrained by data from HERA and the Tevatron. 
In addition to improved
precision on previously measured quantities, new observables with
enhanced or complementary sensitivity to the underlying dynamics
have also been introduced. 

There are fundamental reasons why it is important to understand soft strong interactions
better. It is the non-perturbative 
strong force which binds the fundamental quarks and gluons inside hadrons, 
the detailed mechanisms for which remain poorly understood. In addition
to this question of confinement, soft strong interactions are at the heart of the
dynamical generation of hadronic mass. There may also be deep connections 
through string theory \cite{Aharony:1999ti} between
the soft hadronic degrees of freedom which drive the strong interaction at large
coupling strengths and the weakly coupled regime of gravity. At a more practical
level, it is minimum bias processes which are most pertinent to the modelling
of background activity in the LHC detectors caused by the `pile-up' of multiple
events in the same bunch crossing. Even within a single collision,
the background activity caused by multiple interactions and the underlying event
requires an understanding of minimum bias physics. There are also many
applications beyond the LHC, for example in modelling the development of
cosmic ray air showers \cite{Jung:2009eq}.

\begin{figure}[htb] \unitlength 1mm
  \begin{center}
    \begin{picture}(120,40)
      \put(-18,5){\epsfig{file=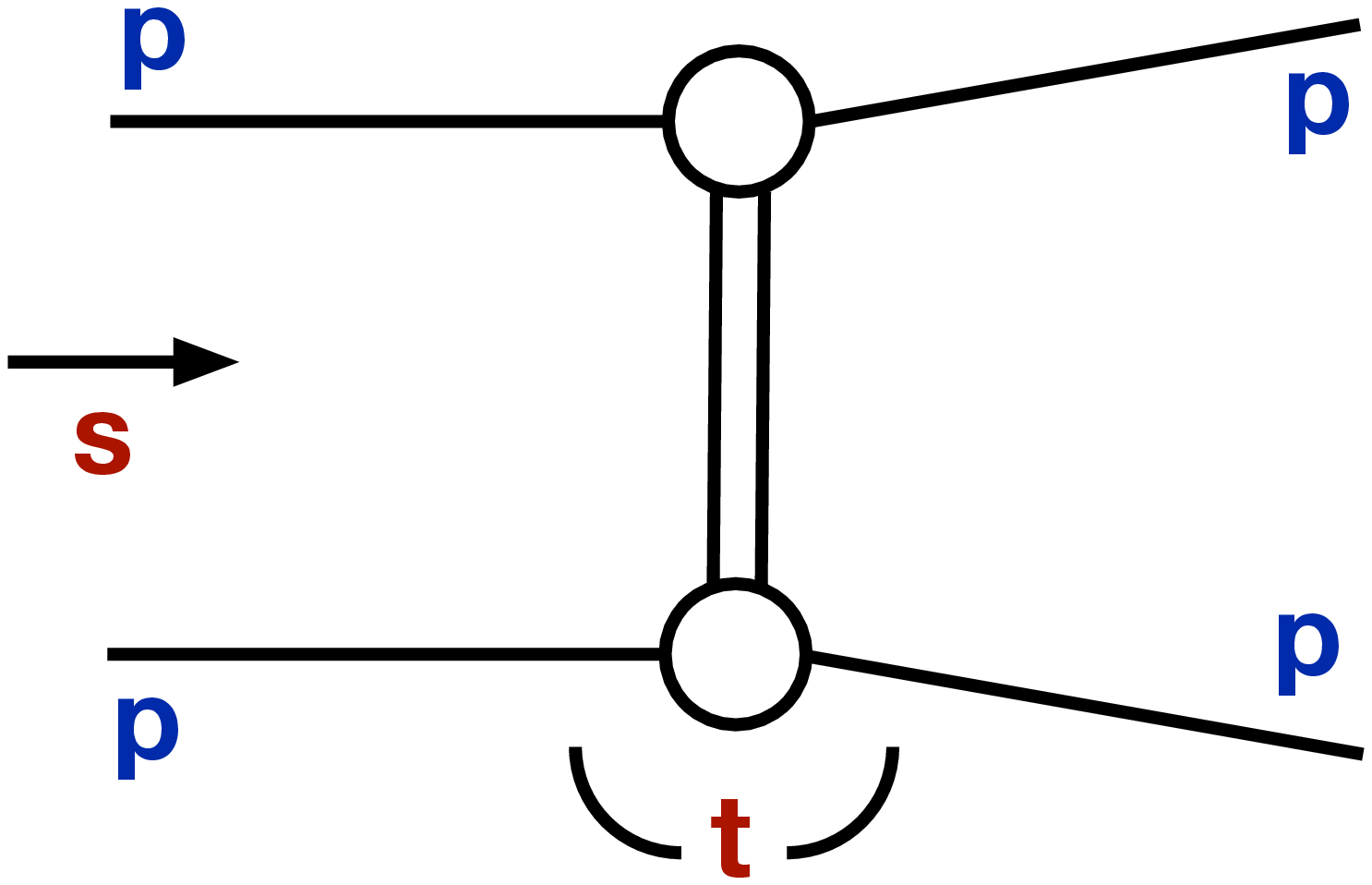,height=0.2\textwidth}}
      \put(37,3){\epsfig{file=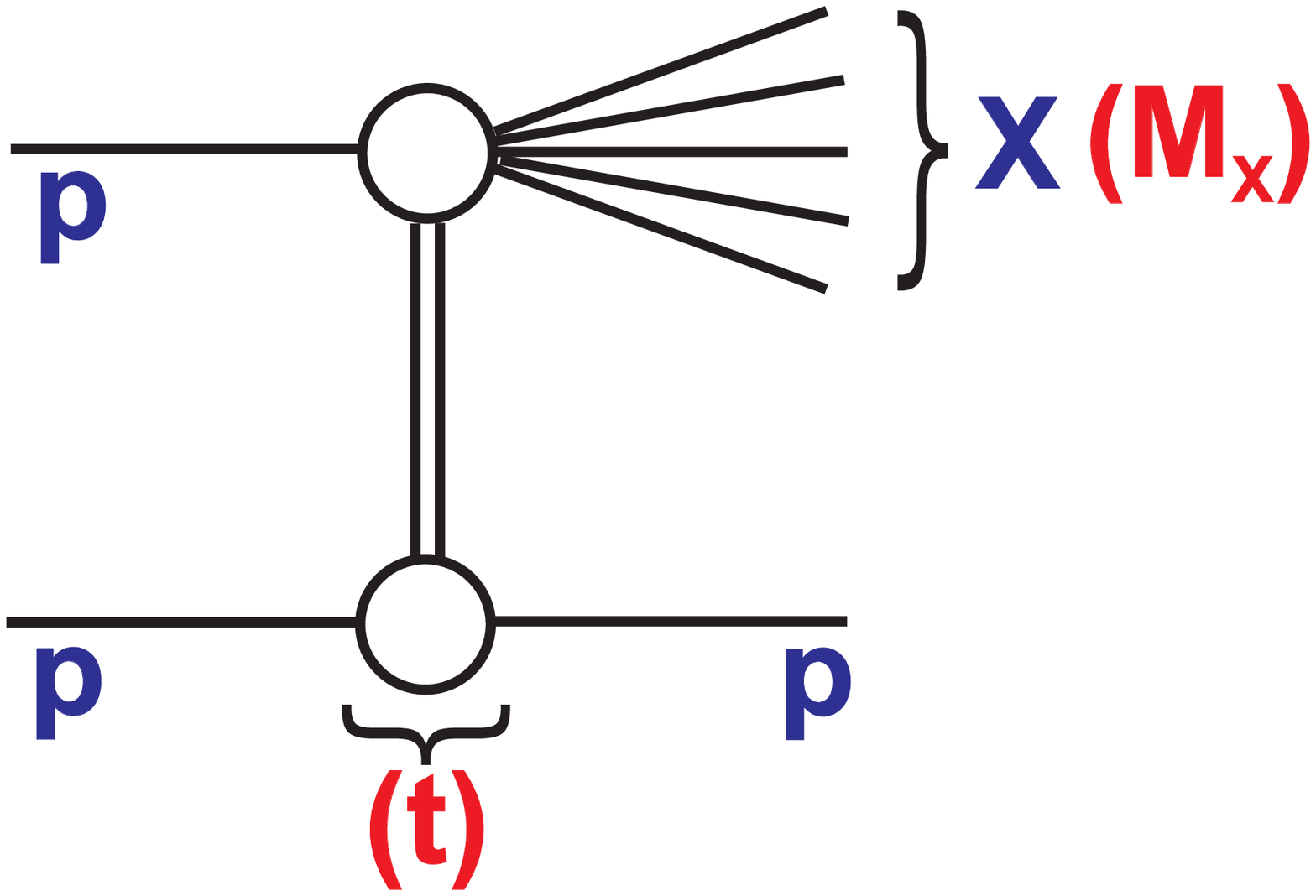,height=0.25\textwidth}}
      \put(95,7){\epsfig{file=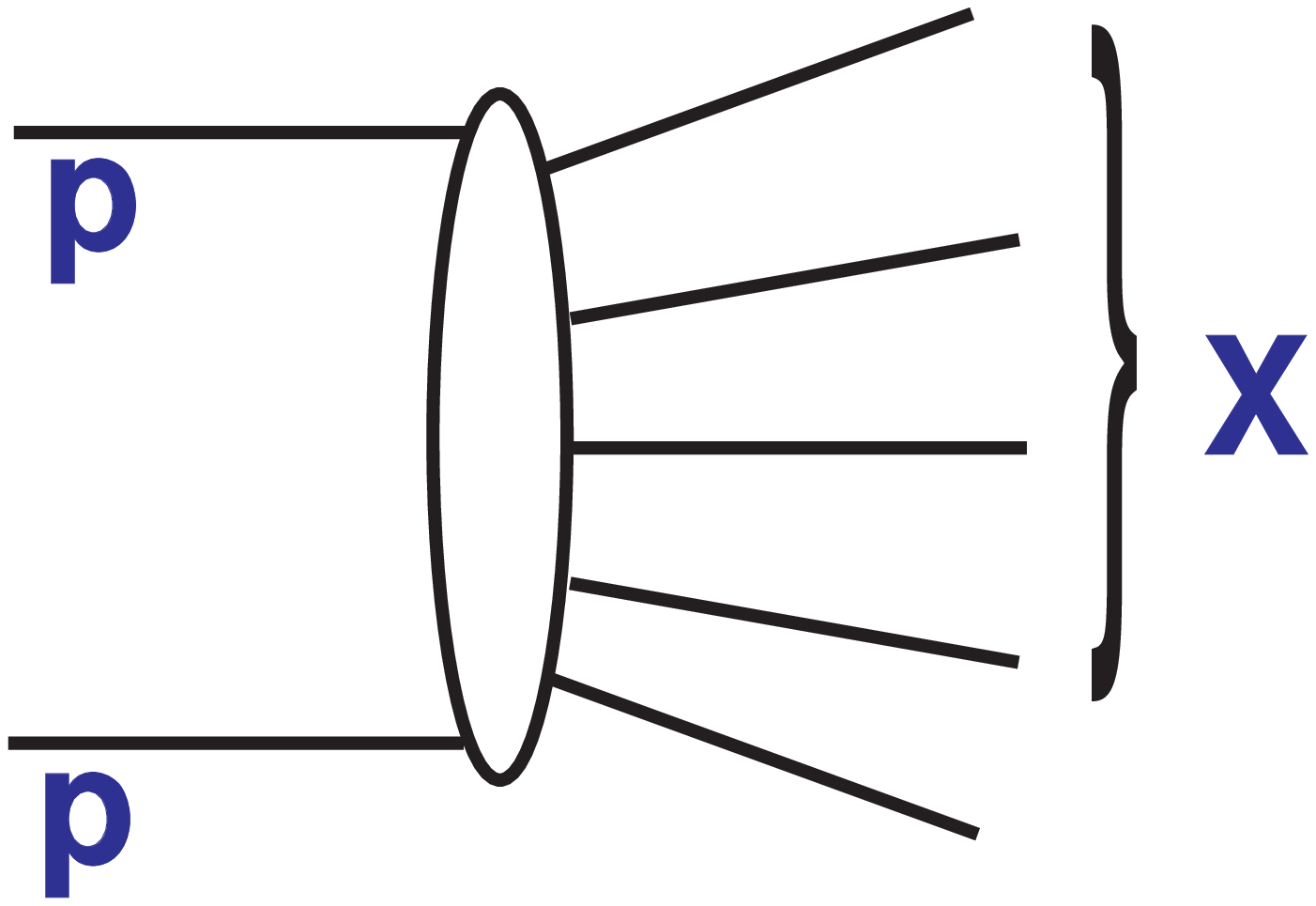,height=0.22\textwidth}}
      \put(3,-3){\Large{\bf{(a)}}}
      \put(50,-3){\Large{\bf{(b)}}}
      \put(100,-3){\Large{\bf{(c)}}}
    \end{picture}
  \end{center}
  \caption{Schematic illustrations of (a) the elastic process
$pp \rightarrow pp$, (b) the single diffractive dissociation process
$pp \rightarrow Xp$ and (c) the non-diffractive process
$pp \rightarrow X$. The kinematic variables discussed in the text
are indicated.}
\label{feynman}
\end{figure}

This talk \cite{slides} covered not only soft interactions, 
but also the interface between the soft and hard regimes,
which is a natural starting point for attempts to make a complete
description of strong interactions.
Hadron-hadron collisions are broken down into
broad categories as illustrated for the case of $pp$
scattering in figure~\ref{feynman}. 
The simplest process is elastic scattering 
($pp \rightarrow pp$, figure~\ref{feynman}a), 
in which the two incoming protons collide at a centre of mass energy
$\surd s$ and remain intact. This process can be described by a single
non-trivial variable, which is usually chosen to be the 
squared four-momentum transfer, $t$. 
Beyond the 
elastic case, the next simplest process is single diffractive
dissociation ($pp \rightarrow Xp$, 
figure~\ref{feynman}b), in which one of the 
beam particles dissociates to produce a
multi-particle excitation $X$, possessing a continuum of masses
$M_X$. In addition to $t$, such processes are described by a 
further invariant, which may be chosen to be either $M_X$, or the
fractional energy loss of the intact proton $\xi = M_X^2 / s$.
In the closely related double diffractive dissociation process ($pp \rightarrow XY$),
both beam particles dissociate to produce two independently
fragmenting and hadronising systems $X$ and $Y$. For LHC
kinematics, the dissociation masses $M_X$ and $M_Y$ may be as
large as $1 \ {\rm TeV}$. The final, non-diffractive,
category ($pp \rightarrow X$, figure~\ref{feynman}c) 
includes all processes not described by the elastic and 
diffractive channels. In the non-diffractive case, particle
production typically takes place throughout the full available 
rapidity region of size $\sim \ln (s / m_p^2)$. Crudely, about half of the total $pp$ 
cross section at LHC energies is attributable to non-diffractive
processes, with the remainder attributable to the
diffractive contributions.

\section{Elastic Scattering and Total Cross Sections}

\begin{figure}[htb] \unitlength 1mm
  \begin{center}
    \begin{picture}(120,140)
      \put(110,110){\Large{\bf{(a)}}}
      \put(110,40){\Large{\bf{(b)}}}
      \put(-7,140){\epsfig{file=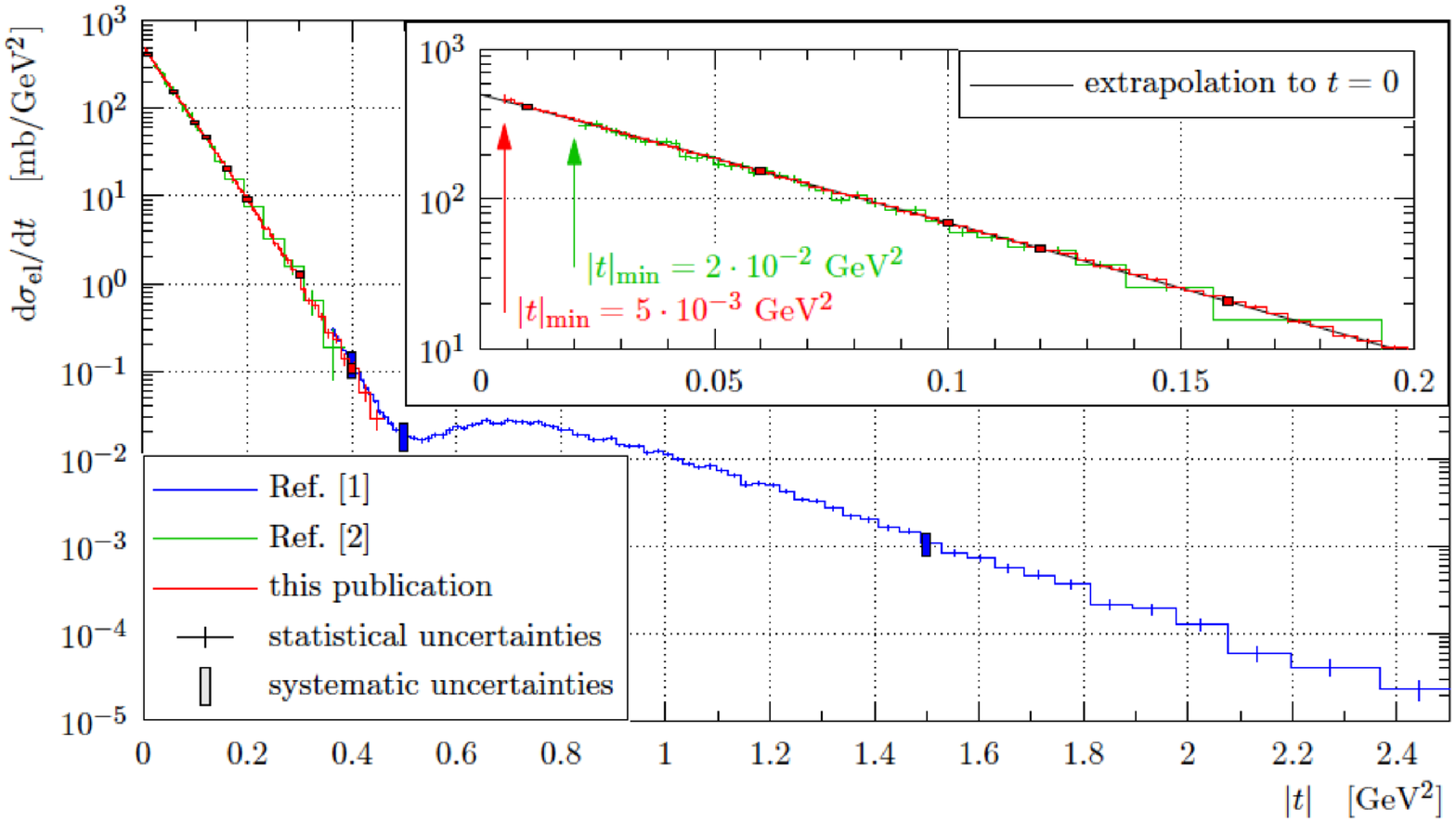,width=0.435\textwidth,angle=270}}
      \put(-5,5){\epsfig{file=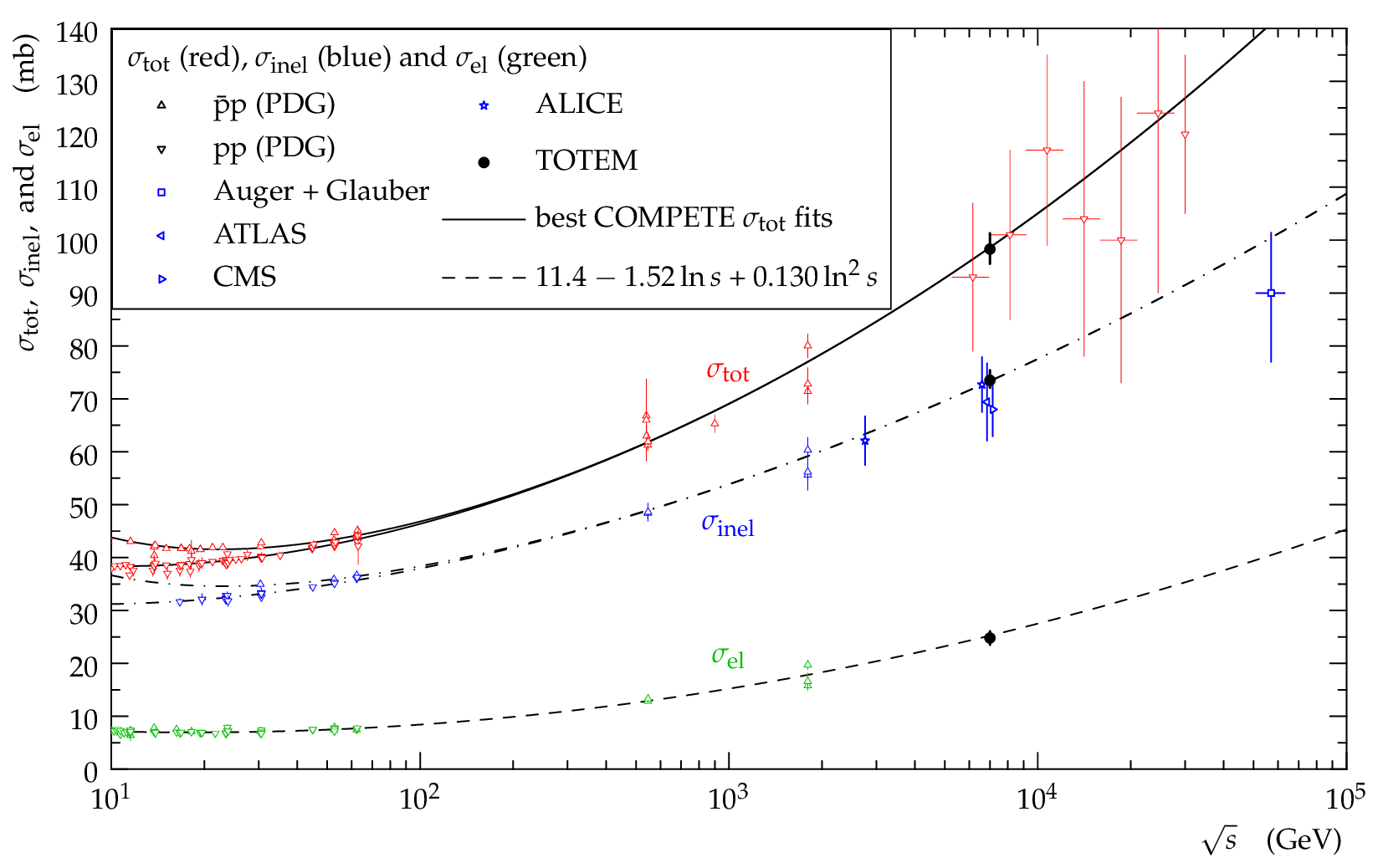,width=0.7\textwidth}}
    \end{picture}
  \end{center}
  \caption{(a) Compilation of TOTEM elastic scattering results \cite{TOTEM:totel}.
The data indicated as `Ref $1$' and `Ref $2$' and the 
curve denoted `this publication' correspond, respectively, to 
citations \cite{Antchev:2011zz}, \cite{Antchev:2011vs} and 
\cite{TOTEM:totel} here.
(b) Compilation of total, total-inelastic and elastic cross section data \cite{Antchev:2011vs}.
Additional results at $\sqrt{s} = 8 \ {\rm TeV}$ have also recently 
appeared \cite{totem:8tev}.}
\label{totem}
\end{figure}

The elastic scattering cross section has recently been measured differentially
in $t$ both by D0 at the 
Tevatron \cite{Abazov:2012qb} and by TOTEM at the 
LHC \cite{TOTEM:totel} (see figure~\ref{totem}a). As has been the case
historically, the low $t$ region is well modelled by an exponential dependence,
${\rm d} \sigma / {\rm d} t \propto \exp{(Bt)}$, such that the slope parameter, $B$
quantifies the spatial extent of the interaction region. The recent
results 
($B = 16.9 \pm 0.2 \ {\rm GeV^{-2}}$ at $\surd s = 1.96 \ {\rm TeV}$ \cite{Abazov:2012qb}
and  
$B = 19.9 \pm 0.3 \ {\rm GeV^{-2}}$ at $\surd s = 7 \ {\rm TeV}$ \cite{TOTEM:totel})
confirm that the slope parameter
increases with energy (shrinkage of the forward elastic peak) and suggest 
a faster growth than was obtained at lower energy and is usually assumed in 
models. A possible interpretation of this lies
in process and energy-dependent absorptive corrections. The increase in $B$
and the decrease of the point in $|t|$ at which
the characteristic minimum in the distribution occurs 
(from around $|t| = 0.7 \ {\rm GeV^{2}}$ at the Tevatron to 
around $0.53 \ {\rm GeV^{2}}$ at the LHC) both indicate that the effective size of the
region over which the protons interact grows with energy, as might be expected as 
longer and longer-lived quantum fluctuations become important. 

It is interesting to compare these results with those from quasi-elastic vector
meson production at HERA ($ep \rightarrow eVp$), where, in the case where a hard scale is provided by
either $Q^2$ or the quark composition of the vector meson, the size of the interaction
region corresponds to the spatial extent of the proton. Recent results, including a 
first measurement of $\Upsilon$ production, have yielded 
an aymptotic limit of $B \sim 5 \ {\rm GeV^{-2}}$ for such processes,
suggesting a rather small effective
proton size of around $0.6 \ {\rm fm}$ \cite{Abramowicz:2011fa}. 
Given that vector meson 
photoproduction is driven at lowest order by two gluon exchange, 
this in turn indicates that the gluon radius of the proton may
be smaller than its quark radius, which is well measured 
with electromagneitc probes
to be around $0.8 \ {\rm fm}$.

Elastic cross sections at $t \rightarrow 0$ 
are closely related to total cross sections, via the
optical theorem. The TOTEM collaboration have exploited this fact to extract the
total $pp$ cross section at LHC energies in several ways, including one method
which is independent of luminosity measurements \cite{totem:8tev}. In 
the example
shown in figure~\ref{totem}b, the total cross section is obtained via
\begin{eqnarray*}
  \sigma_{\rm tot}^2 = \frac{16 \pi}{1 + \rho^2} 
\left( \frac{{\rm d} \sigma_{\rm el}}{{\rm d} t} \right)_{t = 0} \ .
\end{eqnarray*}
The result is consistent with 
either a logarithmic or a power-law rise with $s$ relative to previous data. 
Subtracting the elastic from the total cross section yields the total inelastic
cross section, 
which is also shown in figure~\ref{totem}b. This quantity has been
measured directly by ATLAS \cite{Aad:2011eu}, 
ALICE \cite{:2012sja} and 
CMS \cite{Chatrchyan:2012nj} in addition to the indirect TOTEM
extraction, with consistent results obtained. 

\section{Soft Diffractive Dissociation}

\begin{figure}[htb]
            \hspace*{2.5cm}
            \includegraphics[width=0.575\textwidth]{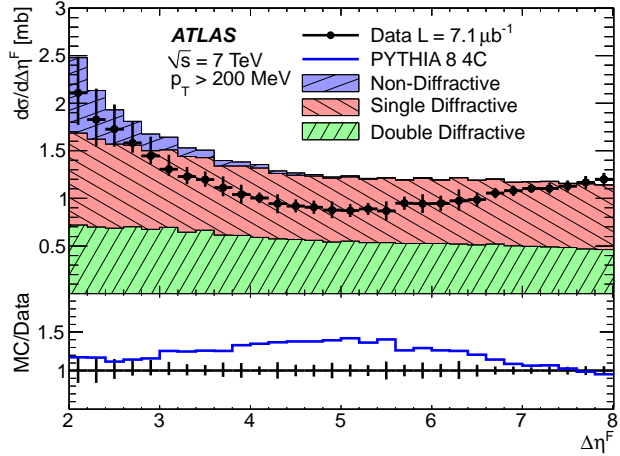}
\caption{Cross section for LHC events containing large rapidity gaps,
extending from at least $\eta = \pm 4.9$ to 
$\eta = \pm 4.9 \mp \Delta \eta_F$ \cite{Aad:2012pw}.
For the contribution from diffractive dissociation processes, the left hand side of the
plot corresponds approximately to $\xi = 10^{-2.5}$ and the right hand side to
$\xi = 10^{-5}$. The data are compared with the predictions of an the PYTHIA8 Monte Carlo
model \cite{Sjostrand:2007gs}, tuned using inclusive ATLAS minimum bias data.}
\label{ATLAS:sdiff}
\end{figure}

Predictions for inelastic diffraction at LHC energies 
varied substantially before the first data were taken, with differences of up to a factor
of two in the predicted cross sections according to the most commonly used 
Monte Carlo models, PYTHIA6 \cite{Sjostrand:2006za}, PYTHIA8 \cite{Sjostrand:2007gs} 
and PHOJET \cite{Engel:1994vs}. 
The first detailed measurement of the dynamics of the process,
by ATLAS \cite{Aad:2012pw},
exploits the close correlation between $\xi$ and the size 
$\Delta \eta$ of the rapidity gap
separating the system $X$ form the intact leading proton
($\Delta \eta \approx - \ln \xi$). Since the coverage of the central components
of the detector is restricted to $|\eta| < 4.9$, the 
chosen experimental observable,
$\Delta \eta^F$, measures the size of the gap relative to $\eta = \pm 4.9$.
At large gap sizes 
$\Delta \eta^F \stackrel{>}{_{\sim}} 2$, 
the differential cross section
${\rm d} \sigma / {\rm d} \Delta \eta^F$ is approximately constant as a function
of gap size (figure~\ref{ATLAS:sdiff}),
as expected where diffractive processes dominate. When viewed in detail, 
the size of the cross section and its residual dependence on $\Delta \eta^F$
are sensitive to the dynamics of soft diffraction, the relative importance of
the single and double diffractive dissociation cross sections and the role of
absorptive corrections \cite{Ryskin:2012az}.

\section{Partonic Structure of Diffractive Dissociation}

\begin{figure}[tb] \unitlength 1mm
  \begin{center}
    \begin{picture}(120,53)
      \put(3,-3){\Large{\bf{(a)}}}
      \put(50,-3){\Large{\bf{(b)}}}
      \put(100,-3){\Large{\bf{(c)}}}
      \put(-10,5){\epsfig{file=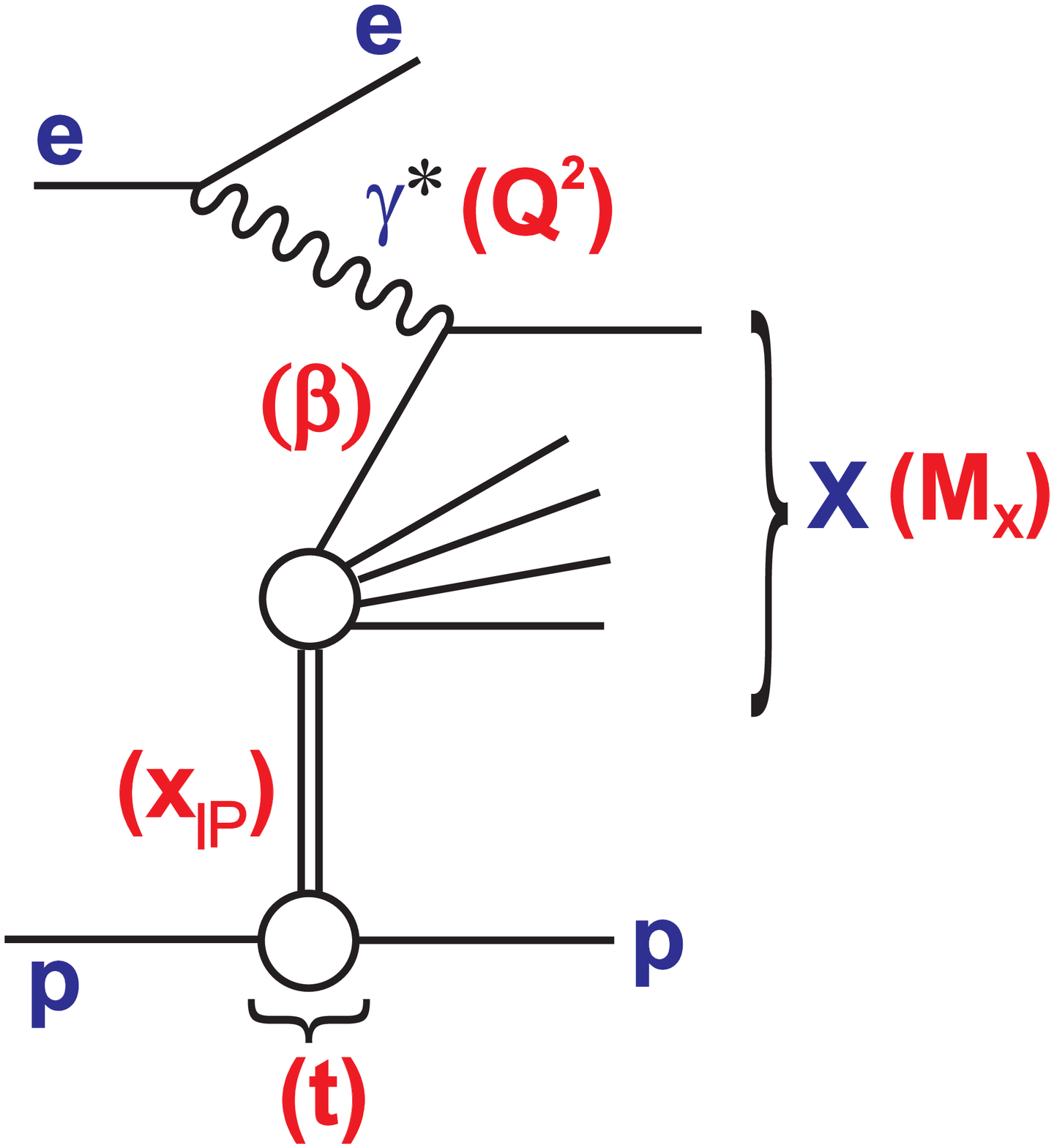,height=0.3\textwidth,clip}}
      \put(42,3){\epsfig{file=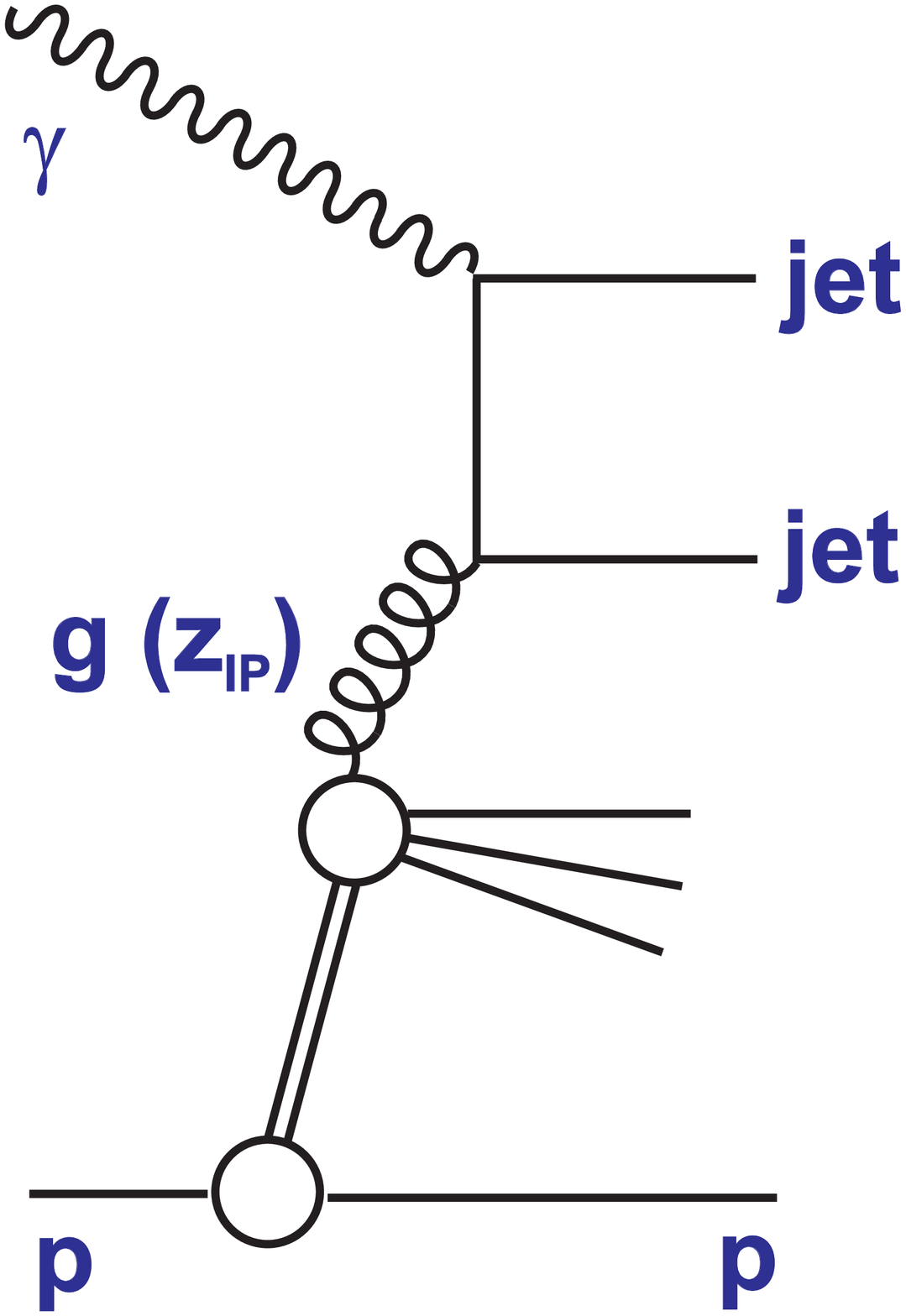,height=0.3\textwidth,clip}}
      \put(90,53){\epsfig{file=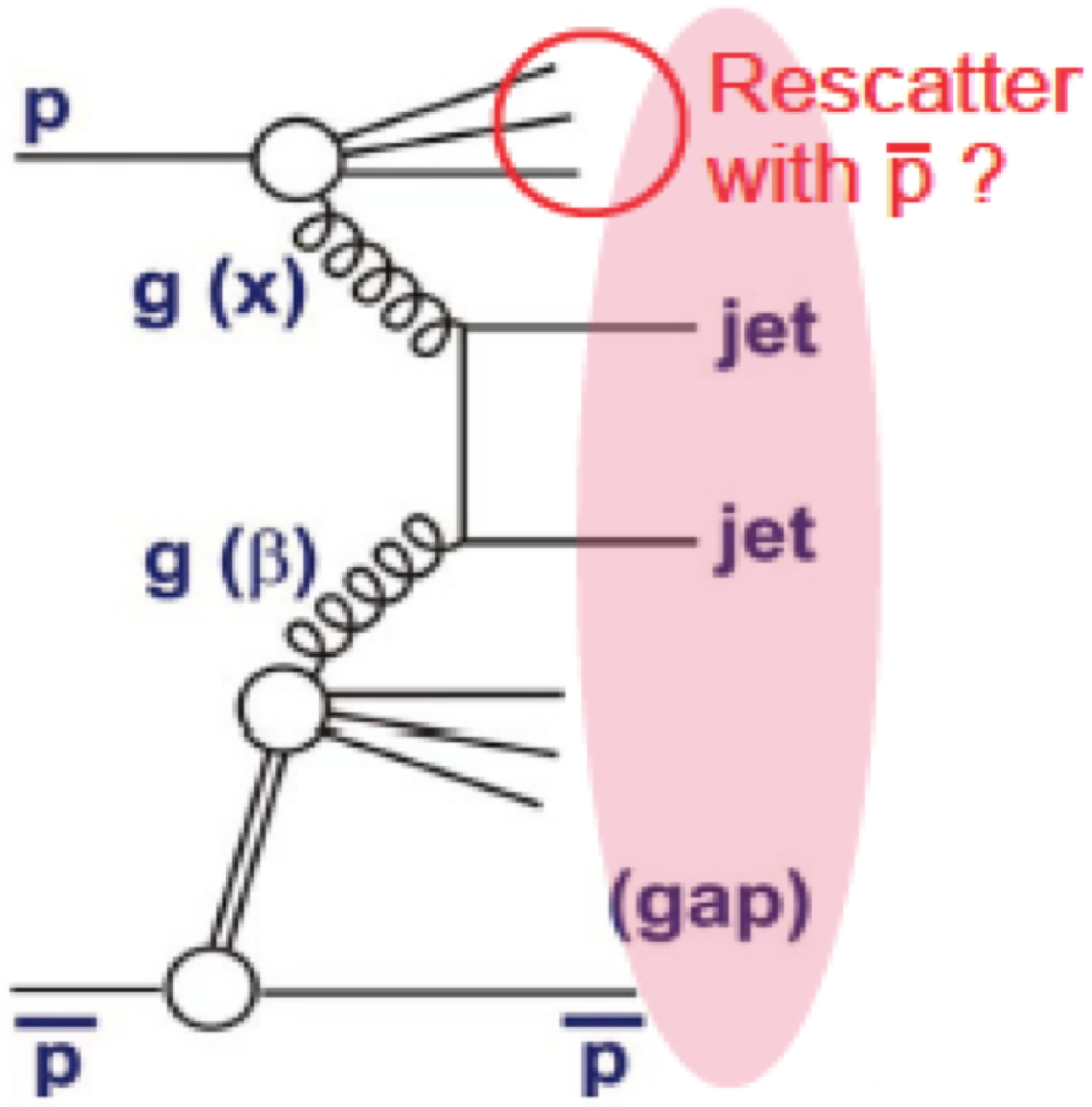,height=0.3\textwidth,angle=270,clip}}
    \end{picture}
  \end{center}
\caption{Sketches of diffractive hard scattering processes.
(a) Inclusive DDIS at the level of the quark parton model,
illustrating the commonly used kinematic variables. 
(b) Dominant lowest order diagram for 
high transverse momentum jet production in 
DDIS, in
which a parton of momentum fraction $z_{I\!\!P}$ from the DPDFs
enters the hard scattering. (c) Diffractive dijet production in $\bar{p} p$
scattering, illustrating the rescattering mechanism which is believed to
produce a rapidity gap survival probability significantly smaller than unity.}
\label{diff:feynman}
\end{figure}

\begin{figure}[htb]
            \hspace*{2.5cm}
            \includegraphics[width=0.525\textwidth]{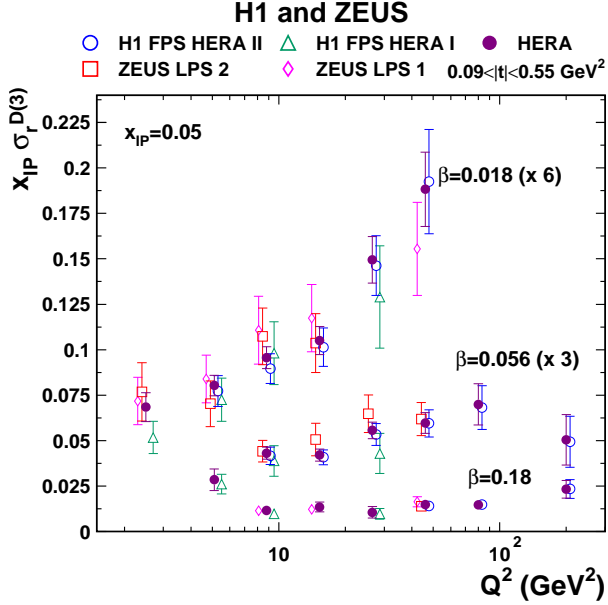}
\caption{$Q^2$ dependence of the DDIS cross section at 
${x_{_{I\!\!P}}} \equiv \xi = 0.05$ for selected values of 
$\beta= x / x_{_{I\!\!P}}$. Results are shown from H1 and ZEUS
measurements in which leading protons are tagged in dedicated
spectrometers well downstream of the interaction point, as well as their
combination \cite{:2012vx}.}
\label{HERA:sdiff}
\end{figure}

The study of shorter distance diffractive processes has been a major theme at the HERA
electron-proton collider, where 
hard scales may be provided 
by the photon virtuality $Q^2$ or large transverse momenta generated in
jets (figures~\ref{diff:feynman}a and~\ref{diff:feynman}b, respectively). 
With the most recent diffractive DIS (DDIS) data, close 
agreement has been reached between 
the H1 and ZEUS measurements. 
A first combination
of inclusive diffractive data from the two experiments, using measurements obtained
by tagging and measuring the leading protons emerging intact from the interactions, 
has recently been published \cite{:2012vx}. The combination is
shown for a selection of illustrative data points in 
figure~\ref{HERA:sdiff}. The improvement in precision is considerably better than
would be expected on statistical grounds alone, due to the effective cross-calibration
between the two measurements implicit in the averaging procedure.

\begin{figure}[htb] \unitlength 1mm
  \begin{center}
    \begin{picture}(120,52)
      \put(-10,-3){\epsfig{file=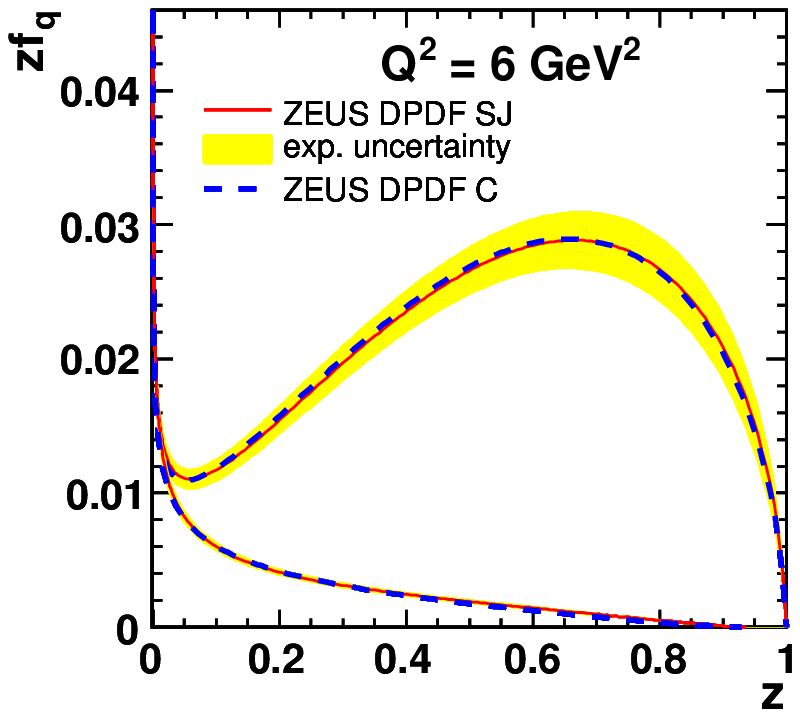,height=0.375\textwidth,clip}}
      \put(60,-3){\epsfig{file=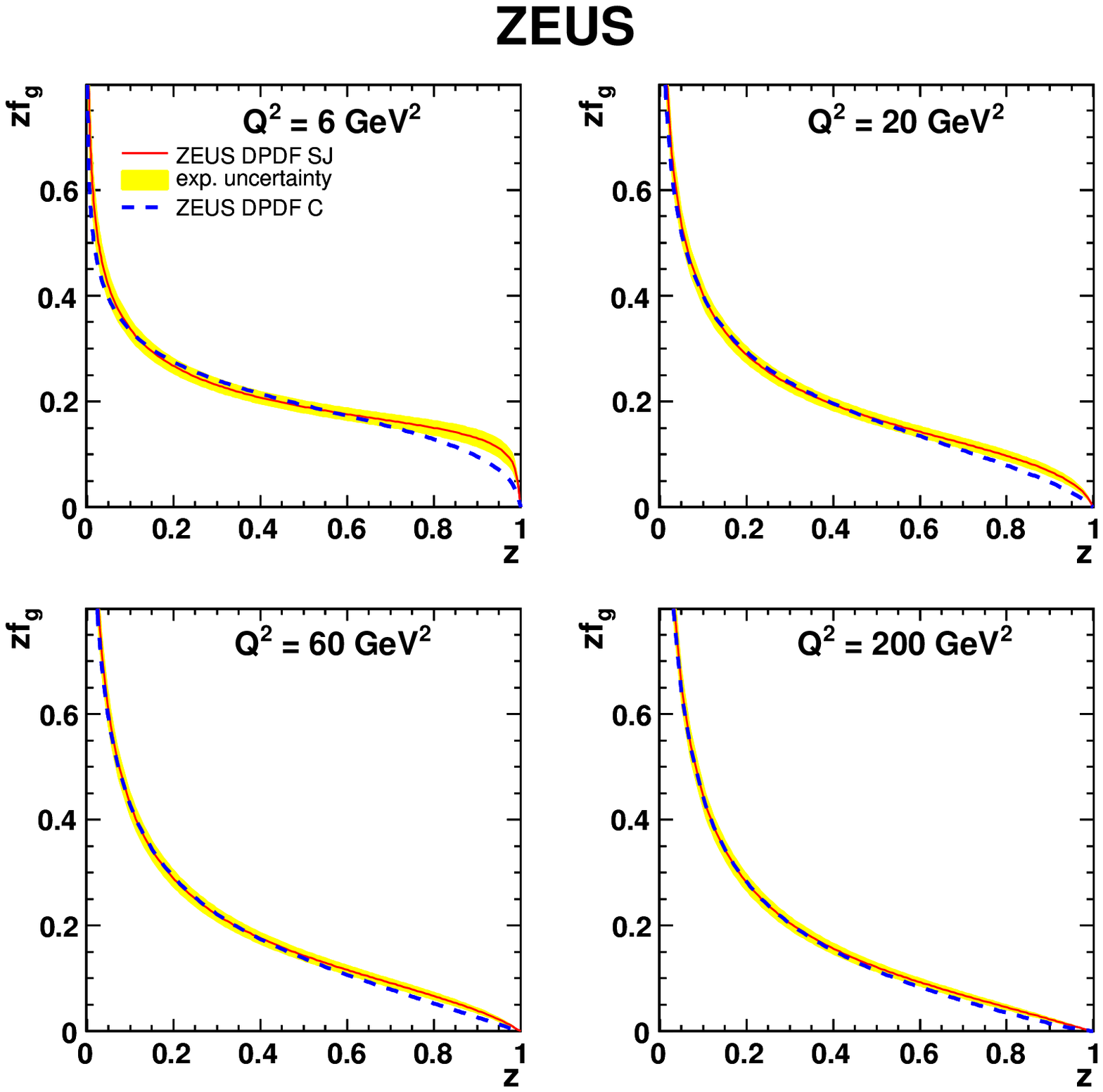,height=0.375\textwidth,clip}}
      \put(24,27){{\Large{light}}}
      \put(12,10){{\Large{charm}}}
    \end{picture}
  \end{center}
  \caption{Diffractive quark (left) and gluon (right) parton distributions
extracted from a ZEUS fit to inclusive DDIS and 
diffractive dijet data \cite{Chekanov:2009aa}.}
\label{HERA:dpdf}
\end{figure}

The inclusive diffractive cross sections, often together with diffractive
dijet data, 
have been subjected to standard next-to-leading-order
QCD fits based on DGLAP evolution, 
to obtain diffractive parton densities (DPDFs), corresponding to conditional
probability distributions for partons of different flavours to be present
at particular fractions of the exchanged momentum (denoted $z$ or $z_{_{I\!\!P}}$) 
under the constraint that the proton
stays intact with a given $\xi$. Recent examples are shown in 
figure~\ref{HERA:dpdf} \cite{Chekanov:2009aa}. These results confirm, with improved
precision, previous conclusions \cite{Aktas:2006hy}
that the DPDFs are dominated by a gluon density,
which extends to large momentum fractions. 

\begin{figure}[htb]
            \hspace*{1cm}
            \includegraphics[width=0.4\textwidth]{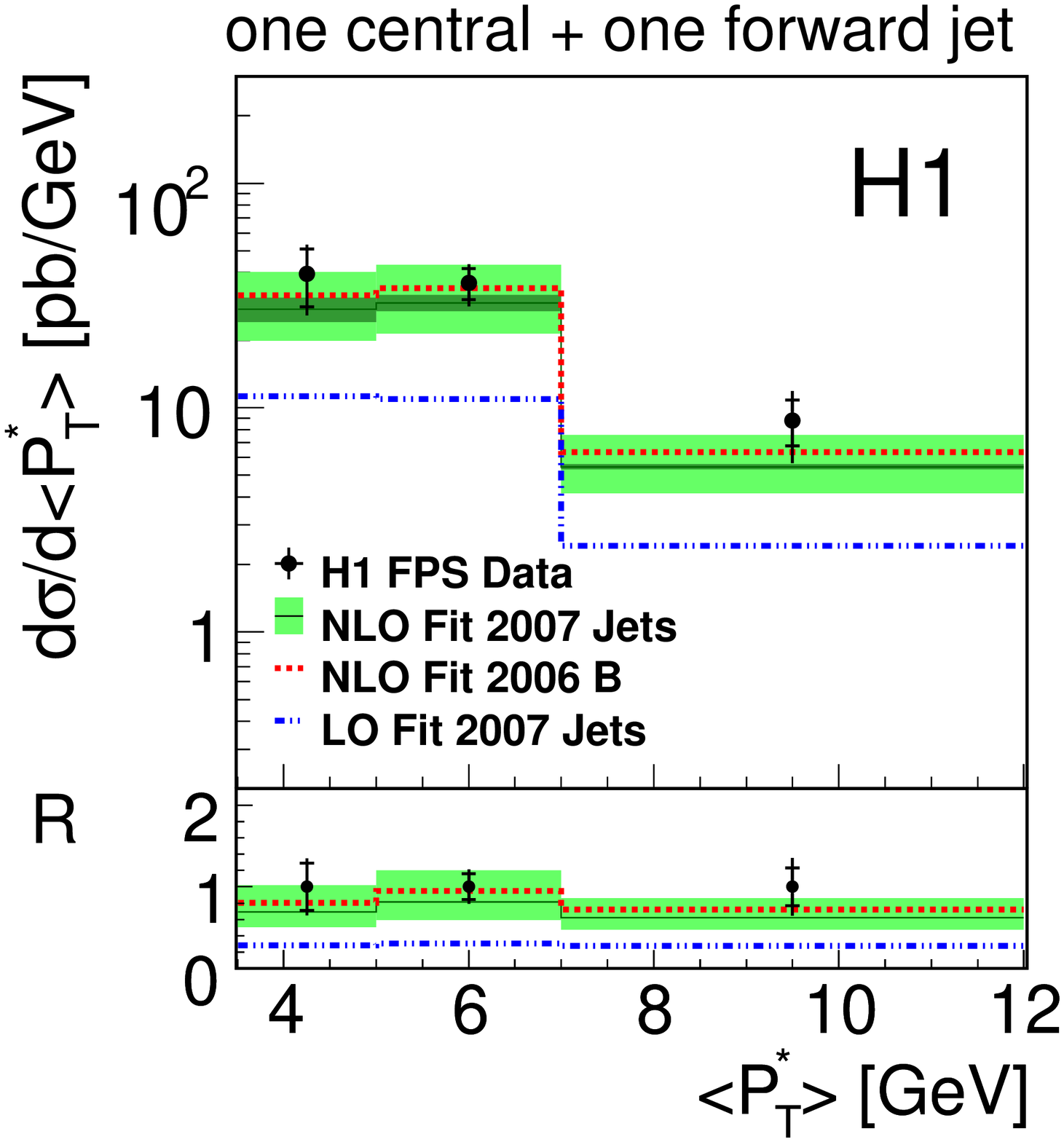}
            \hspace*{0.3cm}
            \includegraphics[width=0.4\textwidth]{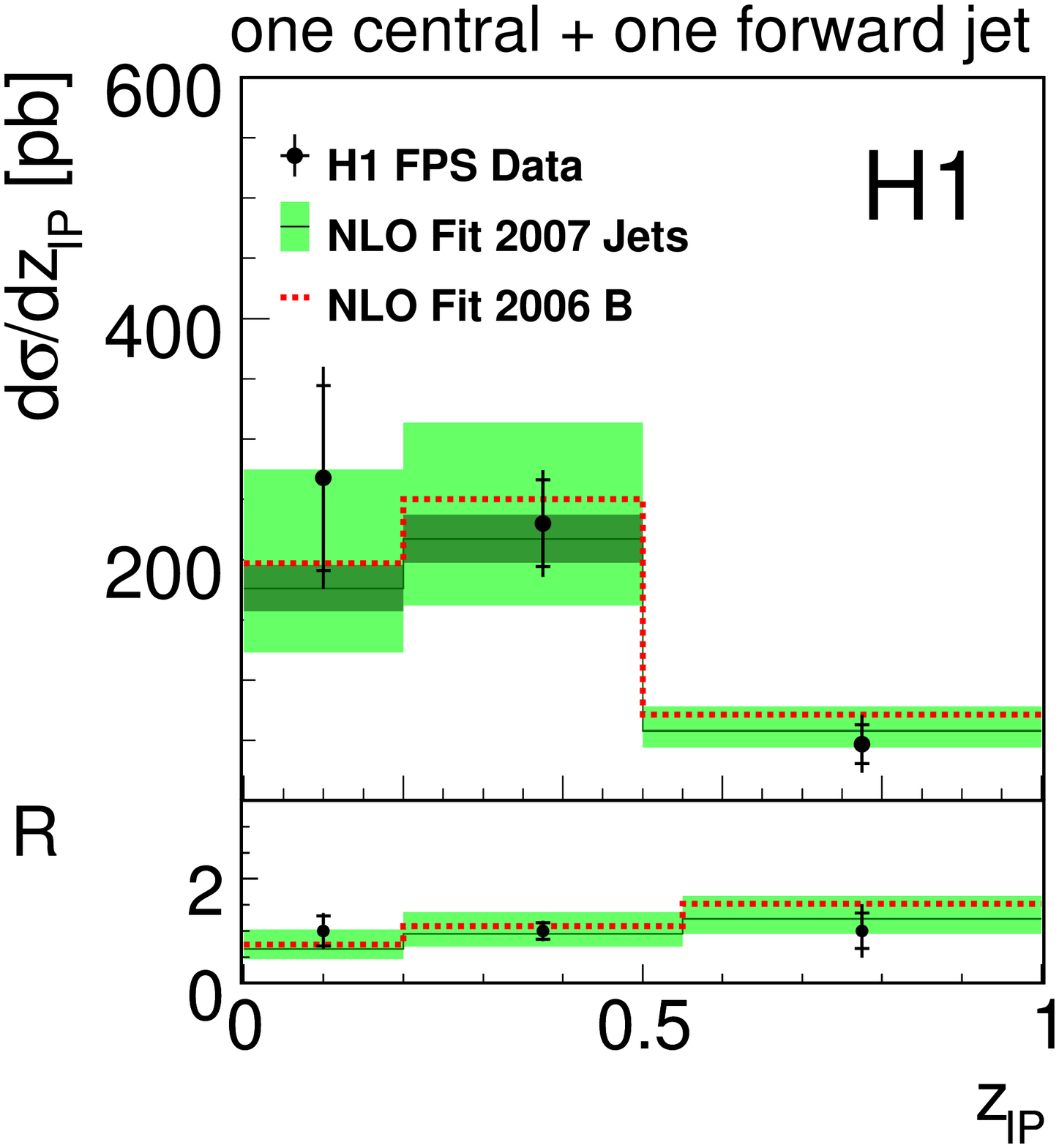}
\caption{Cross sections for the diffractive production of
two jets, at least one of which must be forward of $\eta = 1$ 
in DIS at HERA \cite{Aaron:2011mp}, obtained from events in which an intact
leading proton is detected directly. 
(a) Dependence on the mean jet transverse momentum.
(b) Dependence on a hadron level estimator of the
momentum fraction of the incoming parton from the DPDFs.}
\label{HERA:jets}
\end{figure}

DPDFs such as those shown in figure~\ref{HERA:dpdf} have been applied
as an input to QCD calculations of a wide range of diffractive processes within
DIS. Such comparisons have proved to be universally successful. A recent 
example \cite{Aaron:2011mp} is shown in figure~\ref{HERA:jets}. 
Here, for the first time at HERA, 
dijet cross sections are measured in diffractive
events selected on the basis of an intact reconstructed final state proton.
Using this selection method in place of rapidity gap based techniques 
allows the study of events in which one of the reconstructed jets is close
in rapidity to the edge of the rapidity gap. This topology is
sensitive to possible hard diffractive
production, where the full exchanged momentum enters into the hard subprocess
producing the jets, a process which ought not to be described by the DPDF approach. 
The good agreement in figure~\ref{HERA:jets} indicates that the hard diffractive
process represents only 
a relatively small contribution throughout the accessible kinematic range.  

\begin{figure}[htb]
            \hspace*{2cm}
            \includegraphics[width=0.5\textwidth,angle=270]{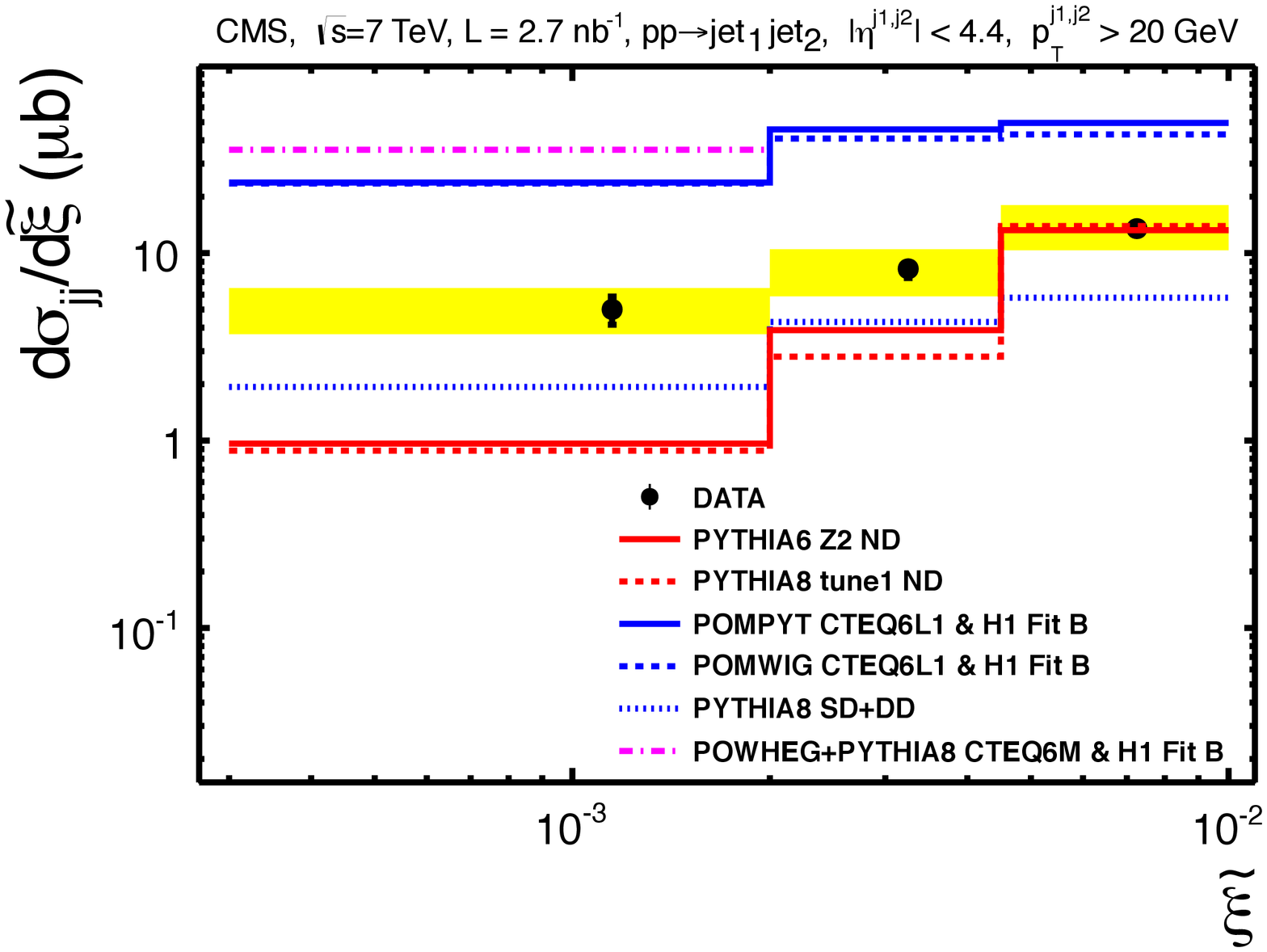}
\caption{CMS Measurement 
of the cross section for the production of pairs of jets in
LHC events containing rapidity gaps (the variable $\tilde{\xi}$ corresponds to
$\xi$ under the interpretation of a diffractive production mechanism). 
Among others, the data are compared
with predictions based on DPDFs from HERA (POMPYT, POMWIG) and with 
models in which
the gaps are produced solely through large fluctuations in the rapidity 
distribution of final state particles in non-diffractive
events (PYTHIA6 Z2, PYTHIA8 tune 1). 
Figure from \cite{Chatrchyan:2012vc}.}
\label{CMS:jets}
\end{figure}

Whilst models based on DPDFs work well to
describe all diffractive dissociation processes 
in DIS, they fail spectacularly 
(by a factor of around 10) when DPDFs extracted from HERA
data are applied to diffractive $\bar{p} p$ scattering at the 
Tevatron \cite{Affolder:2000vb}. This discrepancy
is usually interpreted in terms of
multiple scattering effects, which occur in the
presence of beam remnants (figure~\ref{diff:feynman}c). 
These are closely related to the multiple parton
interactions discussed in section~\ref{nondiff}
and are usually quantified by a `rapidity gap survival
probability'. The gap survival probability is expected to be smaller at the LHC than
at the Tevatron. First LHC results are now 
emerging \cite{Chatrchyan:2012vc}, as shown in figure~\ref{CMS:jets}. 
Despite large uncertainties in the predictions, the data cannot be described
merely by hadronisation fluctuations in non-diffractive processes and models
based on HERA DPDFs without suppression factors clearly overestimate the cross
section. The data have been interpreted in terms of a gap survival probability
of $0.08 \pm 0.04$, which is a little larger than expected. Measurements using
proton-tagged data at the LHC are eagerly awaited in order to suppress the 
poorly understood non-diffractive contributions to such distributions.

\section{Non-Diffractive Processes}
\label{nondiff}

Non-diffractive processes under minimum bias conditions are notoriously
difficult to understand and to model. Couplings are large, such that 
perturbative
tools are not available and
many subtle and inter-connected effects are at play. 
There were many presentations at ICHEP'12 which 
reported measurements of
identified and unidentified particle distributions and energy flows, all of 
which are well suited to the development and tuning of Monte Carlo models which 
aim to give as complete a description as possible of
minimum bias processes. Here a handful of illustrative examples are chosen, 
particularly in areas where 
substantial progress has 
been made recently. 

\begin{figure}[htb]
            \hspace*{2.8cm}
            \includegraphics[width=0.425\textwidth,angle=270]{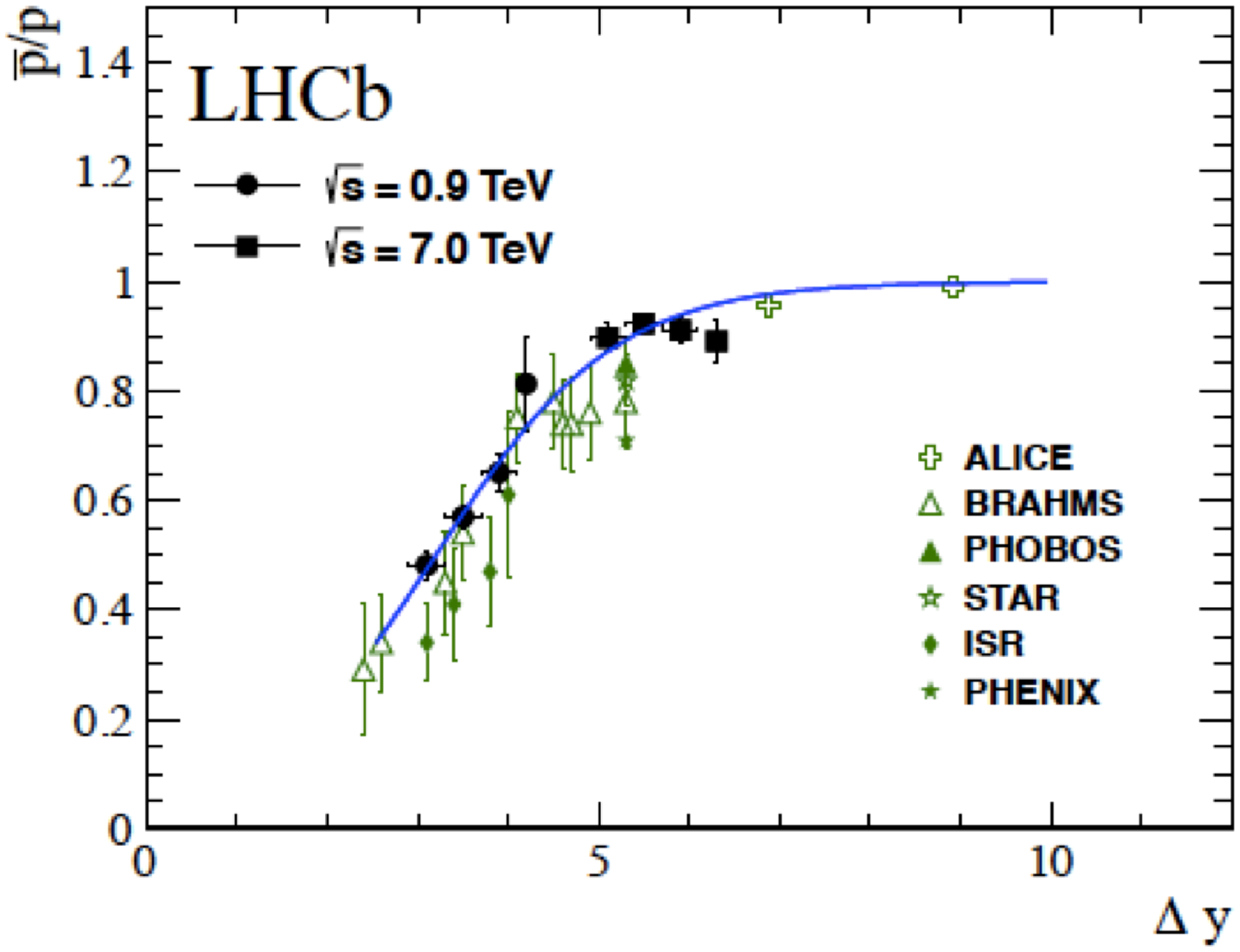}
\caption{Compilation of data on the ratio of production rates of
antiprotons to protons as a function of the rapidity distance 
$\Delta y$ from the outgoing proton beam particles \cite{lhcb:pbar}.}
\label{LHCb:pbarfig}
\end{figure}

The complexity of the minimum bias final state and the long range nature of the 
colour-connections at work are illustrated by a recent ATLAS measurement 
showing a correlation coefficient of larger than 50\% between 
the charged particle multiplicities at $\eta = +2.5$ and 
$\eta = -2.5$ \cite{tim:martin}. Another illustration is shown in 
figure~\ref{LHCb:pbarfig}, in the form of 
measurements by several collaborations
of the ratio of antiproton to proton yields as a function of the rapidity 
distance $\Delta y$ from the outgoing proton beam particles. At very large 
$\Delta y \sim 10$, corresponding to the 
most central region of the detector in the LHC case, 
no influence from the beam 
particle is expected and the proton
and antiproton rates are compatible, yielding a ratio close to unity. However, the ratio falls 
significantly below 1 already for $\Delta y \simeq 5$, indicating that the 
baryon number associated with the beam proton can be transported over 
remarkably large distances.

\begin{figure}[htb]
            \hspace*{2.5cm}
            \includegraphics[width=0.5\textwidth,angle=270,clip]{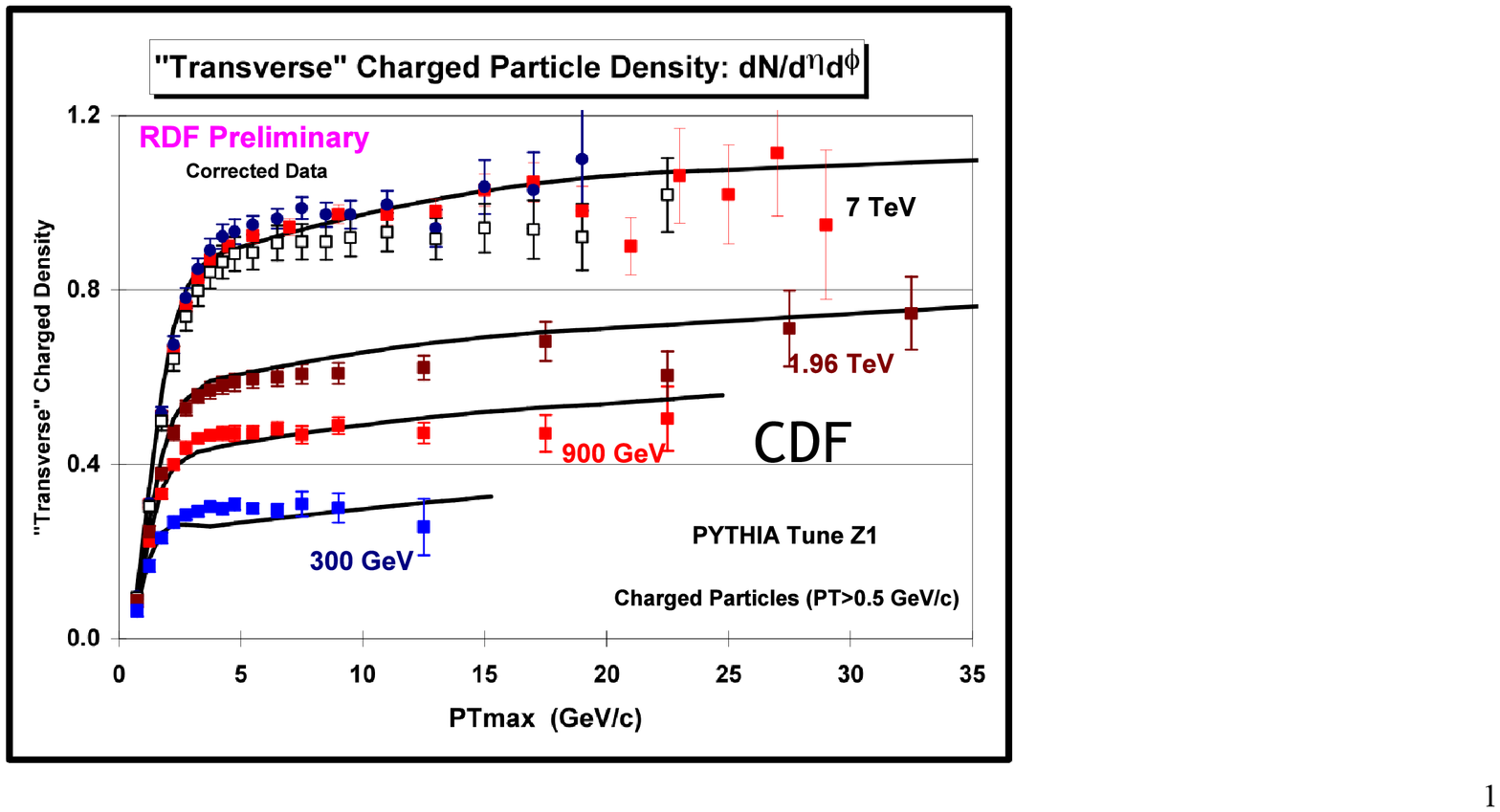}
\caption{Charged particle density,
${\rm d} N / ({\rm d} \eta {\rm d} \phi)$ in a region transverse 
to the leading track in minimum bias samples, as a function of the leading
track transverse momentum, $P_{\rm T, max}$. The $\surd s = 7 \ {\rm TeV}$
data were recorded by ATLAS (blue circles), 
CMS (red squares) and ALICE (open squares).  
Figure from \cite{Rick:Field}.}
\label{Field}
\end{figure}

An area where progress has been impressively rapid since the turn-on of the 
LHC is the understanding of the `underlying event' corresponding to all of the
final state particles produced beyond those associated with the hardest scattering.
The sources of the underlying event are the remnants of the 
colliding hadrons and 
any activity from multiple scatterings in the same event. The latter 
are usually modelled
on the basis of an integral over different impact parameters between 
the two hadrons, for each of which secondary partonic scattering is generated
using proton PDFs at a suitably reduced momentum fraction, 
distributed over an appropriate spatial matter distribution.
Constraints on the  
underlying event have been obtained mainly by studying
charged particle distributions in a region transverse 
in azimuth to the direction of the
hard scatter, as estimated by the direction of the highest transverse momentum
track. The arrival of the first data on such observables from the LHC quickly
showed that the existing models were inadequate, giving rise to an intense
tuning effort. The state of the art in this regard is represented by the 
`Z1' tune of PYTHIA6 and other closely related tunes \cite{Rick:Field}. 
This has been benchmarked using a variety of observables, an example of which is
shown in figure~\ref{Field}. Here, the 
$\eta$ and $\phi$-averaged density of charged particles in the
transverse region is shown as a function of the transverse momentum of the 
leading track. In addition to LHC data at $\surd s = 7 \ {\rm TeV}$, 
data from the Tevatron
at the nominal
$\surd s = 1.96 \ {\rm TeV}$ and from a Tevatron energy scan at
$\surd s = 900 \ {\rm GeV}$ and $\surd s = 300 \ {\rm GeV}$ are also shown. The 
measurements are well described by the Z1 tune throughout. The good description of the
energy dependence bodes well for a high quality early description of LHC data
when the energy increases to $\surd s = 13$ or $14 \ {\rm TeV}$ from 2015. 

There are many other examples of successful 
underlying event predictions by similarly tuned
Monte Carlo models. With increased luminosity, 
these studies have been extended to include a variety of dedicated samples involving
for example dijet or Drell-Yan dimuon production, 
which provide natural hard scales and a clearer direction for the hard 
interaction \cite{gilvan:alves}.
The models also give a good description of alternative 
global event characterisation variables, 
including event shapes
such as thrust \cite{tim:martin}. 
Future progress in this direction is limited by the obvious
complications introduced by high levels of pile-up. 

\begin{figure}[htb]
            \hspace*{2cm}
            \includegraphics[width=0.5\textwidth,angle=270,clip]{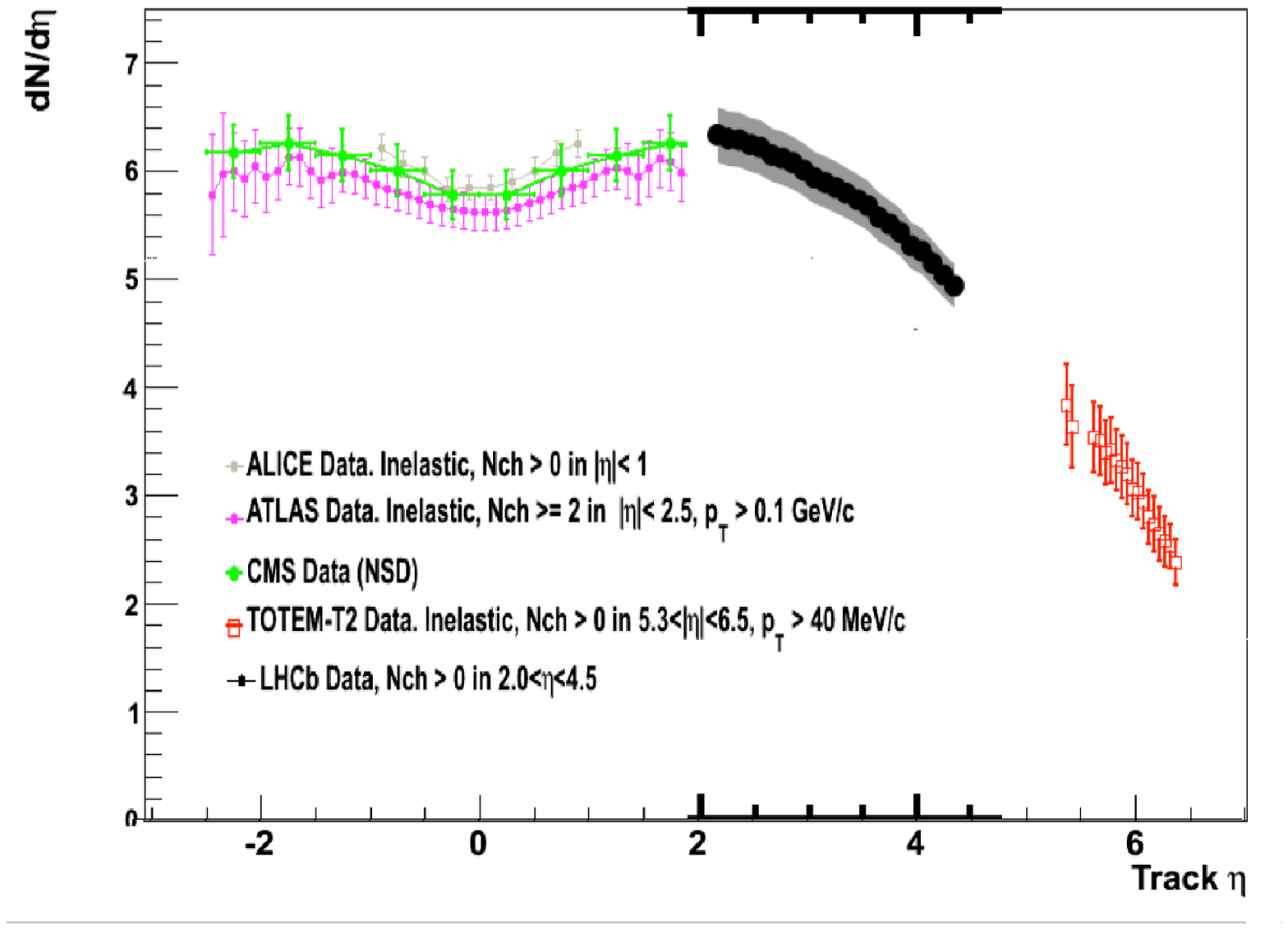}
\caption{Compilation of LHC charged track multiplicity data as a function of 
pseudorapidity \cite{totem:parallel}.}
\label{dNdeta}
\end{figure}

Underlying event studies are usually only sensitive to particle production 
in the
central region, where the best charged particle tracking is located.
Another area where there has been considerable recent 
experimental progress is in measurements
of particle production and energy flow in the forward direction. As illustrated
in figure~\ref{dNdeta}, the LHC experiments between them are sensitive over a 
very wide range, extending well beyond the rapidity plateau, 
with the capability to measure forward charged particle production in addition to 
energy flow. 

\begin{figure}[htb]
            \hspace*{2cm}
            \includegraphics[width=0.5\textwidth,angle=270,clip]{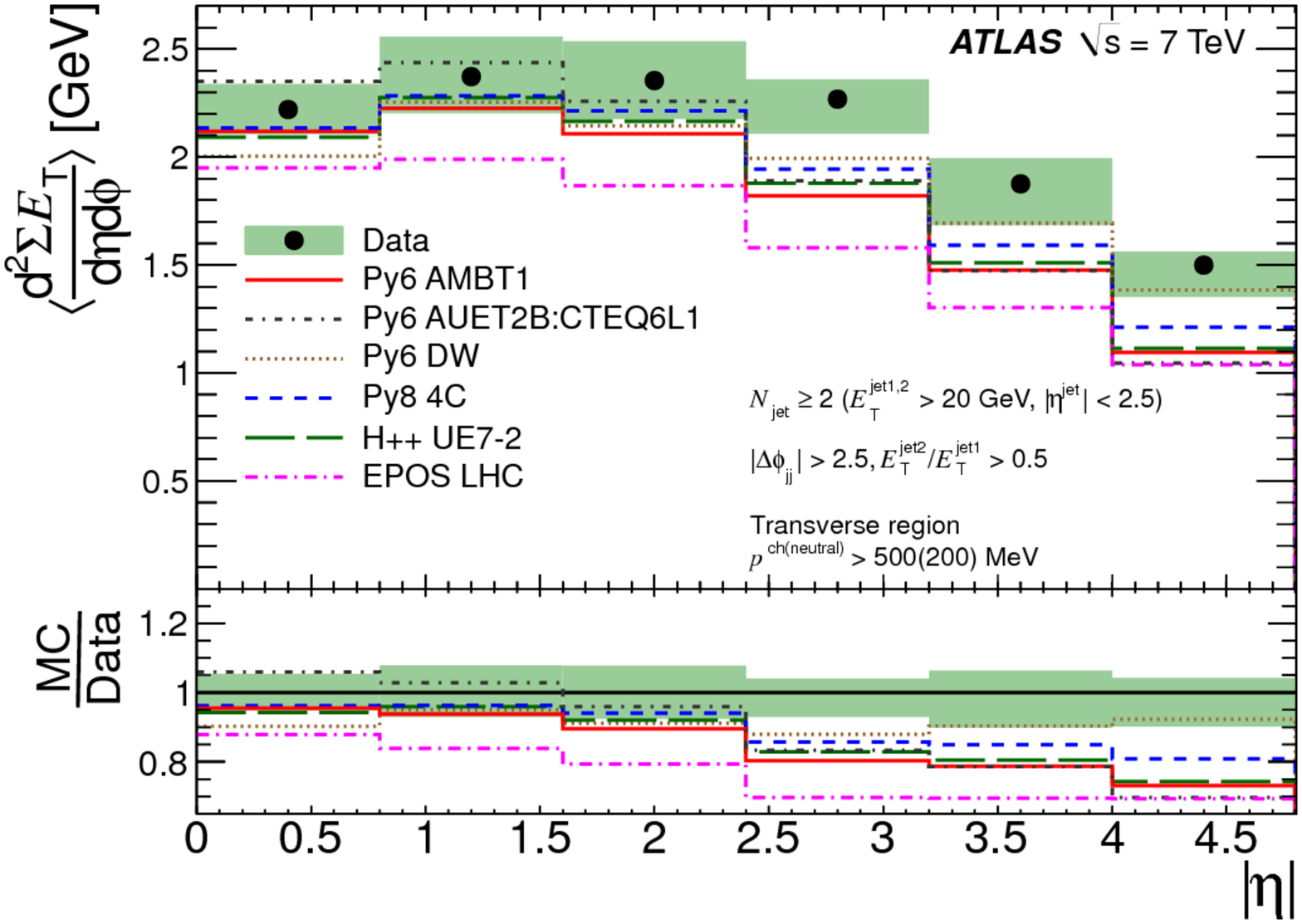}
\caption{ATLAS transverse energy density for minimum bias $\surd s = 7 {\rm TeV}$
LHC $pp$ collisions, averaged over azimuthal angle and the 
indicated intervals of peseudorapidity \cite{:2012dsa}.}
\label{fwd:energy}
\end{figure}

An intense effort has been made to obtain minimum bias data 
extending as forward in rapidity as 
possible \cite{tim:martin,gilvan:alves,totem:parallel,LHCb:parallel}. An 
illustrative
example is shown in figure~\ref{fwd:energy}, where the final state transverse
energy density is measured to $| \eta | = 4.9$. 
Whereas the standard tunes of PYTHIA6 shown here are successful in the central region, 
they lie further and further 
below the data as $|\eta|$ increases. This 
discrepancy may be explained by a combination of
inadequate modelling of the low $x$ gluon density of the proton,
the treatment of diffractive channels, the underlying event 
and parton cascade dynamics. Predictions based on dedicated
models of cosmic air showers such as 
EPOS \cite{Werner:2008zza} are more successful.
Similar conclusions are reached in the other related studies which extend 
even further forward and also address the $\surd s$ 
dependence \cite{CMS:fwd,Aspell:2012ux,Chatrchyan:2011wm,Aaij:2011yj}.

\section{Summary}

The introduction of new data in a previously unexplored energy regime at the LHC
has given new motivations and a renewed stimulus for the reinvestigation of the
softest processes taking place in hadronic scattering. In addition to
the impressive precision obtained, the introduction of new 
observables and the investigation of extended 
kinematic regions has given complementary 
insights. Together with continuing high
quality measurements from HERA and the Tevatron, this has led to a deeper understanding
of long-standing problems such as the underlying dynamics of multi-parton interactions and 
of diffractive processes, as well
as increasingly reliable tools for a complete model of hadronic physics at the TeV
scale. 

\section*{Acknowledgements}

It was a great privilege to attend the conference and to give a plenary talk - 
sincere thanks
to the organisers for the invitation. For help in preparing the talk or for
supplying material, I would also like to thank G.~Alves, S.~Bhadra, R.~Cieselki,
M.~Diele, R.~Field, C.~Glasman, A.~Grebenyuk, 
M.~Klein, T.~Martin, R.~Muresan, H.~Niewiadoniski,
R.~Polifka, D.~Salek, A.~Soffer and V.~Simak.


\begin{thebibliography}{99}

\bibitem{waltzing} Banjo Paterson {\em `Waltzing Matilda'} (1895).

\bibitem{Aharony:1999ti}
  O.~Aharony,
  Phys.\ Rept.\  {\bf 323} (2000) 183
  [hep-th/9905111].

\bibitem{Jung:2009eq}
  Z.~Ajaltouni {\it et al.},
{\em `Proceedings of HERA and the LHC workshop series'}, chapter 5
  [arXiv:0903.3861 [hep-ph]].

\bibitem{slides} Slides available at 
\verb+epweb2.ph.bham.ac.uk/user/newman/diffraction/ICHEP12-Plenary.pdf+

\bibitem{Abazov:2012qb}
  D0 Collaboration,
  Phys.\ Rev.\ D {\bf 86} (2012) 012009
  [arXiv:1206.0687 [hep-ex]].

\bibitem{TOTEM:totel}
TOTEM Collaboration, {\em `Measurement of proton-proton elastic scattering and total cross-section at $\surd s = 7 TeV$'}, CERN-PH-EP-2012-239.

\bibitem{Antchev:2011zz}
  TOTEM Collaboration,
  Europhys.\ Lett.\  {\bf 95} (2011) 41001
  [arXiv:1110.1385 [hep-ex]].

\bibitem{Antchev:2011vs}
  TOTEM Collaboration,
  Europhys.\ Lett.\  {\bf 96} (2011) 21002
  [arXiv:1110.1395 [hep-ex]].

\bibitem{totem:8tev}
TOTEM Collaboration, {\em `A luminosity-independent measurement 
of the proton-proton total cross-section at $\surd s = 8 \ {\rm TeV}$'}, 
CERN-PH-EP-2012-354.

\bibitem{Abramowicz:2011fa}
  ZEUS Collaboration,
  Phys.\ Lett.\ B {\bf 708} (2012) 14
  [arXiv:1111.2133 [hep-ex]].


\bibitem{Aad:2011eu}
  ATLAS Collaboration,
  Nature Commun.\  {\bf 2} (2011) 463
  [arXiv:1104.0326 [hep-ex]].

\bibitem{:2012sja}
  ALICE Collaboration,
  [arXiv:1208.4968 [hep-ex]].

\bibitem{Chatrchyan:2012nj}
  CMS Collaboration,
[arXiv:1210.6718 [hep-ex]].

\bibitem{Sjostrand:2006za}
  T.~Sjostrand, S.~Mrenna and P.~Skands,
  JHEP {\bf 0605} (2006) 026
  [hep-ph/0603175].

\bibitem{Sjostrand:2007gs}
  T.~Sjostrand, S.~Mrenna and P.~Skands,
  Comput.\ Phys.\ Commun.\  {\bf 178} (2008) 852
  [arXiv:0710.3820 [hep-ph]].

\bibitem{Engel:1994vs}
  R.~Engel,
  Z.\ Phys.\ C {\bf 66} (1995) 203.

\bibitem{Aad:2012pw}
  ATLAS Collaboration,
  Eur.\ Phys.\ J.\ C {\bf 72} (2012) 1926
  [arXiv:1201.2808 [hep-ex]].

\bibitem{Ryskin:2012az}
  M.~Ryskin, A.~Martin and V.~Khoze,
  Eur.\ Phys.\ J.\ C {\bf 72} (2012) 1937
  [arXiv:1201.6298 [hep-ph]].

\bibitem{:2012vx}
  H1 and ZEUS Collaborations,
  Eur.\ Phys.\ J.\ C {\bf 72} (2012) 2175
  [arXiv:1207.4864 [hep-ex]].

\bibitem{Chekanov:2009aa}
  ZEUS Collaboration,
  Nucl.\ Phys.\ B {\bf 831} (2010) 1
  [arXiv:0911.4119 [hep-ex]].

\bibitem{Aktas:2006hy}
  H1 Collaboration,
  Eur.\ Phys.\ J.\ C {\bf 48} (2006) 715
  [hep-ex/0606004].

\bibitem{Aaron:2011mp}
  H1 Collaboration,
  Eur.\ Phys.\ J.\ C {\bf 72} (2012) 1970
  [arXiv:1111.0584 [hep-ex]].

\bibitem{Affolder:2000vb}
  CDF Collaboration,
  Phys.\ Rev.\ Lett.\  {\bf 84} (2000) 5043.

\bibitem{Chatrchyan:2012vc}
  CMS Collaboration,
  Phys.\ Rev.\ D {\bf 87} (2013) 012006
  [arXiv:1209.1805 [hep-ex]].

\bibitem{lhcb:pbar}
LHCb Collaboration, {\em `Measurement of the 
$\bar{p} / p$ ratio in LHCb at $\surd s =$ 900 GeV and 7 TeV'},
LHCb-CONF-2010-009.

\bibitem{tim:martin} T.~Martin, {\em `ATLAS studies of diffraction, soft particle production and double parton scattering'}, parallel session talk at this conference. 

\bibitem{Rick:Field} R.~Field, {\em `The Energy Dependence of the Underlying Event in Hadron-Hadron Collisions'}, parallel session talk at this conference. 

\bibitem{gilvan:alves} G.~Alves, {\em `Forward Physics Results from CMS'},
parallel session talk at this conference.

\bibitem{totem:parallel} H.~Niewiadomski {\em `New measurements of Forward Physics in the TOTEM Experiment at the LHC'}, 
parallel session talk at this conference.  

\bibitem{:2012dsa}
  ATLAS Collaboration,
  JHEP {\bf 1211} (2012) 033
  [arXiv:1208.6256 [hep-ex]].

\bibitem{LHCb:parallel} R.~Muresan {\em `Studies of soft QCD at LHCb'}, 
parallel session talk at this conference.  

\bibitem{Werner:2008zza}
  K.~Werner,
  Nucl.\ Phys.\ Proc.\ Suppl.\  {\bf 175-176} (2008) 81.

\bibitem{CMS:fwd}
  CMS Collaboration,
  arXiv:1302.2394 [hep-ex].

\bibitem{Aspell:2012ux}
  TOTEM Collaboration,
  Europhys.\ Lett.\  {\bf 98} (2012) 31002
  [arXiv:1205.4105 [hep-ex]].

\bibitem{Chatrchyan:2011wm}
  CMS Collaboration,
  JHEP {\bf 1111} (2011) 148
   [Erratum-ibid.\  {\bf 1202} (2012) 055]
  [arXiv:1110.0211 [hep-ex]].

\bibitem{Aaij:2011yj}
  LHCb Collaboration,
  Eur.\ Phys.\ J.\ C {\bf 72} (2012) 1947
  [arXiv:1112.4592 [hep-ex]].

\end{thebibliography}
\end{document}